# Generation of relativistic positrons carrying intrinsic orbital angular momentum

Shaohu Lei,[1,2] Zhigang Bu,[1,*] Weiqing Wang,[1,2] Baifei Shen[1,3,4] and Liangliang Ji,[1,3,†]

[1]*State Key Laboratory of High Field Laser Physics, Shanghai Institute of Optics and Fine Mechanics, Chinese Academy of Sciences, Shanghai 201800, China*

[2] *University of Chinese Academy of Sciences, Beijing 100049, China*

[3]*CAS Center for Excellence in Ultra-intense Laser Science, Shanghai 201800, China*

[4]*Shanghai Normal University, Shanghai 200234, China*

**Abstract:** High energy positrons can be efficiently created through high-energy photons splitting into electron-positron pairs under the influence of the Coulomb field. Here we show that a new degree of freedom-the intrinsic orbital angular momentum (OAM) can be introduced into relativistic positrons when the incident photons are twisted. We developed the full-twisted scattering theory to describe the transfer of angular momentum before and after the interaction. It is found that the total angular momentum (TAM) of the photon is equally distributed among the positron and electron. For each photon TAM value, the generated leptons gain higher average OAM number when the photon spin is anti-parallel to its TAM. The impact of photon polarization on the OAM spectrum profile and the scattering probability is more significant at small photon TAM numbers, owing to the various interaction channels influenced by flipping the photon spin. Our work provides the theoretical basis to study OAM physics in particle scattering and to obtain copious relativistic vortex positrons through the Beth-Heitler process.

\* zhigang.bu@siom.ac.cn

† jill@siom.ac.cn

**I. Introduction:**

High energy positrons are of great significance in modern particle physics experiments. Their collision with energetic electrons is essential in generating new particles such as B-mesons/Z-bosons [1,2] and monitoring various reaction processes via Bhabha scattering [3–5]. Further, in astrophysics positrons are strongly correlated to black hole physics [6], gamma-ray bursts [7] and pair plasma physics [8,9]. Here interactions mainly concern the energy/momentum and the spin properties of the involved positrons. Recently, it is pointed out that particles can also carry intrinsic orbital angular momentum (OAM) as it does for optical photons [10–12], in the presence of vortex states [13–18]. In transmission electron microscope (TEM), vortex electrons have been prepared in the $80 \sim 300$ keV regime to improve the resolution and reveal new information of the subjects [19]. Interactions based on relativistic vortex particles have also been studied in the framework of quantum electrodynamics (QED), showing new features in the Vavilov–Cherenkov radiation [20,21] and the possibility of creating spin polarized particles from spinless ones [22].

Although high energy vortex particles bring novel insights in many interactions, generation of relativistic vortex positrons is extremely challenge. First, positrons, unlike electrons, must be created before introducing any features to the particle states. Second, manipulation of the lepton wave packets in TEM becomes invalid in the relativistic regime since typical wavelengths of high-energy leptons are too small. Recalling that vortex gamma-photons can be readily obtained from Compton backscattering of Laguerre-Gaussian laser beams off high energy electrons [23], we propose a scheme to generate vortex positrons at MeV energies by bombarding the vortex gamma-photons onto high-Z material. Relativistic electron-positron pairs can be created through the Beth-Heitler (BH) process, which facilitates efficient positron sources. By introducing a new degree of freedom-OAM into the interaction, we obtain the law of vorticity transfer from the incident gamma-photons to the created pairs. This is achieved with the first full vortex scattering theory of the BH process developed in this work.

## II. Theory of the vortex BH process

The natural unit system $\hbar = c = 1$ is applied in all calculations. We consider the process shown in Fig. 1(a). Here a high-energy vortex photon bombards a high-Z target. Under the influence of the Coulomb field, photons may split into electron-positron pairs carrying the initial OAM information. To capture the vortex nature of the interaction, we use the Bessel modes to describe all involved particles [18,23] in the cylindrical momentum space $\bm{k}' = (k'_\perp, \phi_{k'}, k'_z)$, with OAM defined with respect to the same common z-axis. It is a superposition of plane wave states with the same energy but different phase $\phi_{k'}$. The twisted photon takes the form $A^{l,\lambda,\mu}_{k_\perp,k_z}(x) = \int \tilde{A}_l(\bm{k}'_\perp) A^{\lambda,\mu}_{\bm{k}'_\perp,k_z}(x) k'_\perp dk'_\perp d\phi_{k'}$, where $l$ is total angular momentum (TAM) and $\lambda = \pm 1$ is the polarization parameter. Here $A^{\lambda,\mu}_{\bm{k}}(x) = 1/\sqrt{(2\pi)^3 (2\omega)} \varepsilon^{\lambda,\mu}_{\bm{k}} e^{-ik\cdot x}$ is the plane wave state and $\tilde{A}_l(\bm{k}'_\perp) = 1/(\sqrt{2\pi} i^l k'_\perp) \delta(k'_\perp - k_\perp) e^{il\phi_{k'}}$ is the Fourier spectrum, with $\varepsilon^{\lambda}_{\bm{k}}$ the polarization vector ($(\varepsilon^{\lambda}_{\bm{k}} \cdot k = 0)$). The twisted photon field can be expressed as:

$$A_{k_\perp,k_z}^{l,\lambda;\mu}(x) = \varepsilon_{k_\perp,k_z}^{l,\lambda;\mu}(r)e^{ik_z z - i\omega t}$$

$$= \frac{e^{ik_z z - i\omega t}}{4\pi\sqrt{\omega}} \begin{pmatrix} 0 \\ (i/2)\left[(1-k_z/\omega)\Theta_{k_\perp}^{l+\lambda}(r) + (1+k_z/\omega)\Theta_{k_\perp}^{l-\lambda}(r)\right] \\ (\lambda/2)\left[(1-k_z/\omega)\Theta_{k_\perp}^{l+\lambda}(r) - (1+k_z/\omega)\Theta_{k_\perp}^{l-\lambda}(r)\right] \\ (\lambda k_\perp/\omega)\Theta_{k_\perp}^{l}(r) \end{pmatrix}, \qquad (1)$$

where $\omega$ is the photon energy, $\Theta_{k_\perp}^{n}(r) = J_n(k_\perp r)e^{in\theta}$ is the transverse function, and $J_n(r)$ is the Bessel function of the first kind. The Coulomb field takes the form $A_0^{Coul}(\vec{x}) = -Ze/|\vec{x}| = -4\pi Ze\int 1/(2\pi)^3 e^{-i\vec{q}\cdot\vec{x}}/|\vec{q}|^2 d^3q$. We notice that the vortex photon field is not the eigenmodes of the OAM operator $\hat{L}_z = -i\partial/\partial\theta$ defined along the z-direction. For weak spin-orbital coupling, the OAM value can be well represented by the quantity $l$-$\lambda$.

The twisted electron and positron states can be constructed from the positive and negative-frequency plane wave solutions of the Dirac equation $\psi_p^{+,s}(x) = 1/\sqrt{2(2\pi)^3}\, e^{-ip\cdot x}\left(\sqrt{1+M/E}\,\xi^s, \sqrt{1-M/E}(\sigma\cdot\kappa)\xi^s\right)^T$ and $\psi_p^{-,s}(x) = 1/\sqrt{2(2\pi)^3}\, e^{ip\cdot x}\left(\sqrt{1-M/E}(\sigma\cdot\kappa)\eta^s, \sqrt{1+M/E}\,\eta^s\right)^T$, where $\kappa = p/|p|$ is the unit vector of the momentum, $\xi^s$ and $\eta^s$ are the two-component spinors characterizing the electron and positron spins in the rest frame. The Fourier spectrum is the same as that of photons. The one for positron is therefore [18]

$$\psi_{p_\perp,p_z}^{-,m,s}(x) = v_{p_\perp,p_z}^{m,s}(r,\theta)e^{-ip_z z + iEt} = \frac{e^{-ip_z z + iEt}}{\sqrt{2}(2\pi)}\left[\begin{pmatrix}\sqrt{1-\frac{M}{E}\frac{p_z}{|p|}}\sigma^3\eta^s \\ \sqrt{1+\frac{M}{E}}\eta^s\end{pmatrix}\Theta_{p_\perp}^{m}(r) - \frac{ip_\perp}{|p|}\sqrt{1-\frac{M}{E}}\begin{pmatrix}\sigma_{m,p_\perp}^{\perp}(r,\theta)\eta^s \\ 0\end{pmatrix}\right],$$

(2)

with $\sigma_{p_\perp}^{\perp,m}(r,\theta) = \begin{pmatrix} 0 & -\Theta_{p_\perp}^{m-1}(r) \\ \Theta_{p_\perp}^{m+1}(r) & 0 \end{pmatrix}$. The wave function of the twisted electron is similar with a different plane wave bispinor $u_{p_\perp,p_z}^{m,s}(r,\theta)$. Here $m$ stands for OAM number and $s$ the spin number. In principle, the OAM number is not the eigenvalue of the vortex wave function. It is however a good approximation if the spin-orbial interaction (SOI) is not significant. This is the case for the parameters considred here, hence we take $m$ as the OAM in the following and discuss the effect of SOI in later sections. In the perturbation theory, the scattering matrix of twisted photons splitting into positron-electron pairs is $S_{fi} = S_1 + S_2$. The first term is :

$$S_1 = -\frac{ie^2}{2}\int d^4x d^4x' \frac{d^4q}{(2\pi)^4} \bar{\psi}_{p_{1\perp},p_{1z}}^{+,m_1,s_1}(x) \slashed{A}_{k_\perp,k_z}^{l,\lambda}(x) \frac{(\slashed{q}+M)}{(q^2-M^2)} e^{-iq\cdot(x-x')} \gamma^0 \psi_{p_{2\perp},p_{2z}}^{-,m_2,s_2}(x') A_0^{Coul}(x').$$

(3)

Here $M$ is the mass of electron/positron and $\slashed{A} = A_\mu \gamma^\mu$. Substituting the twisted photon state (1), the Coulomb field and electron-positron states (2) into equation (3), we get

$$S_1 = \frac{iZe^3}{4(2\pi)}\sqrt{\frac{(E_1-M)(E_2-M)}{\omega E_1 E_2}} \frac{1}{|\mathbf{p}_1||\mathbf{p}_2|} \delta(\omega - E_1 - E_2) \xi^{s_1\dagger} \Xi_{k_\perp,k_z}^{l,\lambda}(m_1,p_{1\perp},p_{1z};m_2,p_{2\perp},p_{2z}) \eta^{s_2} \Big|_{\substack{E_q=-E_2=E_1-\omega \\ q_z=p_{1z}-k_z}}$$

(4)

with

$$\Xi_{k_\perp,k_z}^{l,\lambda}(m_1,p_{1\perp},p_{1z};m_2,p_{2\perp},p_{2z}) = \begin{pmatrix} \varsigma_{11}\delta_{l,m_1-m_2} & \varsigma_{12}\delta_{l,m_1-m_2+1} \\ \varsigma_{21}\delta_{l,m_1-m_2-1} & \varsigma_{22}\delta_{l,m_1-m_2} \end{pmatrix}.$$

(5)

The integer $l$ represents the photon TAM number, $m_1$ and $m_2$ are the OAM numbers of electrons and positrons, respectively. Details of the matrix elements $\varsigma_{11}$, $\varsigma_{12}$, $\varsigma_{21}$, $\varsigma_{22}$ are included in Appendix. The matrix $S_2$ is obtained by exchanging the photon and the

Coulomb field in $S_1$, which leads to a different matrix $\tilde{\Xi}$ (see in Appendix). Each matrix element in $\Xi$ and $\tilde{\Xi}$ determines the creation probability of pairs with different spin-polarizations, and the angular momentum (AM) dependent Kronecker delta function gives the corresponding selection rule for the twisted BH process,

$$l = m_1 - m_2 + \Delta, \qquad (6)$$

with $\Delta=0, \pm 1$. The minus sign before $m_2$ in Eq. (6) is consistent with the definition of positron AM. To calculate the creation probability, we use the wave-packet to describe the incident photon $\mathcal{A}_\mu^{\lambda,l}(x) = \int 1/\sqrt{2\omega}\, \rho(k_\perp, k_z) A_{k_\perp,k_z;\mu}^{\lambda,l}(x) dk_z dk_\perp$, where the weighting function taking Gaussian distributions $\rho(k_\perp, k_z) = N_{\tau_\perp,\tau_z} \exp[-(k_\perp - \tilde{k}_\perp)^2/\tau_\perp^2 - (k_z - \tilde{k}_z)^2/\tau_z^2]$, with central momenta and energy $\tilde{k}_\perp, \tilde{k}_z$ and $\tilde{\omega}$. The wave packet widths in transverse and longitudinal directions are $\tau_\perp$ and $\tau_z$. Under the narrow wave packet approximation $\tau_{\perp(z)} \ll \tilde{k}_{\perp(z)}$, we get

$$\left|S_{wave-packet}\right|^2 = C \times Z^2 \pi \alpha^3 \delta(\tilde{\omega} - E_1 - E_2) \frac{(E_1 - M)(E_2 - M)}{\tilde{\omega} E_1 E_2 |\mathbf{p}_1|^2 |\mathbf{p}_2|^2} \mathrm{Tr}\left[\xi^{s_1} \xi^{s_1\dagger} (\Xi + \tilde{\Xi}) \eta^{s_2} \eta^{s_2\dagger} (\Xi + \tilde{\Xi})^\dagger\right],$$

(7)

where

$$C = \frac{N_{\tau_\perp,\tau_z}^2}{2} \exp\left(-\frac{2\tilde{\omega}^2}{\tau^2}\right) \frac{\sqrt{\pi}\tau}{2\sqrt{2}} \int d\theta_{k'} d\theta_k \sin\theta_{k'} \sin\theta_k$$
$$\times \left[1 + \mathrm{Erf}\left(\frac{1}{\sqrt{2}\tau}\left(\tilde{k}_\perp(\sin\theta_{k'} + \sin\theta_k) + \tilde{k}_z(\cos\theta_{k'} + \cos\theta_k)\right)\right)\right] \qquad (8)$$
$$\times \exp\left[\frac{1}{2\tau^2}\left(\tilde{k}_\perp(\sin\theta_{k'} + \sin\theta_k) + \tilde{k}_z(\cos\theta_{k'} + \cos\theta_k)\right)^2\right] \tilde{\omega}^3$$

is a coefficient associated with wave packet and $\theta$ is the opening angle of the incident particle in the wave packet $\tan(\theta) \equiv k_\perp/k_z$. The pair creation probability is

$$d\mathcal{P} = p_{1\perp} p_{2\perp} \left| S_{wave-packet} \right|^2 dp_{1\perp} dp_{1z} dp_{2\perp} dp_{2z}. \tag{9}$$

### III. Results

In order to calculate the scattering probabilities, we set the central energy and momenta of the photon wave packet as $\tilde{\omega} = 5\text{MeV}$, $\tilde{k}_z = 4\text{MeV}$, $\tilde{k}_\perp = 3\text{MeV}$, the photon polarization is $\lambda = 1$ and TAM number is $l = 6$. We consider photons interacting with copper atoms $(Z = 29)$. By integrating the momentum of the electron in Eq. (9) we obtain the creation probability of the positron. As known in plane-wave scattering, the interaction follows energy and momentum conservation, leading to a thin resonance line in the momentum space, as represented by the dashed circle in Fig. 1(b). However, in vortex scattering the conservation is for energy. The resonance condition is significantly relaxed such that probability is distributed in a broad region in Fig. 1(b). Furthermore, it is seen that the probability peaks along a certain angle. In Fig. 1(c) we summarize the angular-dependent distribution of positrons at different OAM number $m_2$. While the highest value varies with the OAM, all profiles exhibit peaks at similar angles. The average angular distribution is around $\theta \approx 0.68$, which coincides with the azimuthal angle of the incident photon $\tan(\theta) = k_\perp / k_z = 3/4, \theta = 0.64$. In other words, azimuthal angle is largely conserved in paraxial collision, at which most positrons are created. One should notice that in planewave scattering the scattering angle is dependent on the out-going particle energy.

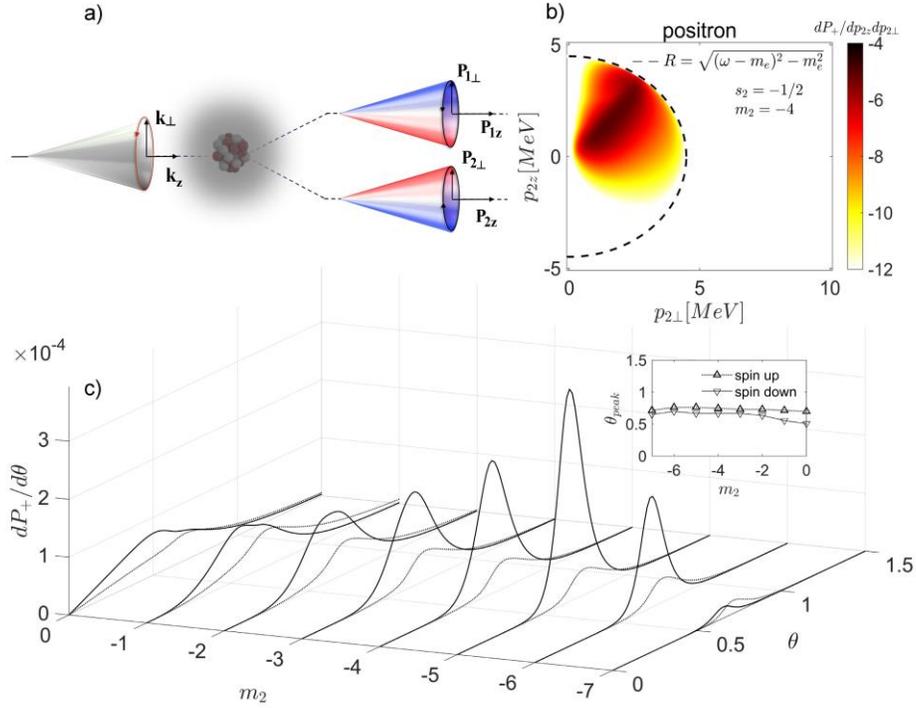

**Figure 1.** (a) Schematic diagram of the twisted Beth-Heitler process where vortex electron-positron pair is created from vortex photons bombarding high-z atoms. (b) The probability distribution of the created positron of $m_2 = -4, s_2 = -1/2$ in the momentum domain. (c) The probability distribution as a function of the opening angle at different OAM values for $s_2 = 1/2$ (black solid) and $s_2 = -1/2$ (black dashed). Parameters of the incident photon are $l = 6, \lambda = 1, \tilde{\omega} = 5 \text{MeV}$.

We also compare the distributions of final positron with spin-up and down states in Fig. 1(c). For the case we considered, the spin-down channel contributes to the large part of the interaction probability. Thus, the generated particle is obviously polarized. This is consistent with the propagation of polarization from polarized photons to positrons [24–26]. The polarization is however not 100%, indicating spin-orbital coupling during the scattering process.

A central question about the vortex scattering is how OAM is distributed among the final particles. In the following, we show OAM spectra for the generated pair and their relationship with the TAM of the initial photon. We integrate the momentum and sum over the spins of created pairs to get the total probabilities at different OAM number, as shown in Fig. 2(a) with photon TAM $l=5,10,15,20$. The electron primarily carries OAM with the same sign as the photon while that of the positron is opposite. This is because the defined direction of positron OAM is opposite to that of electron. As the photon TAM increases, the central OAM shifts towards large values accordingly. Furthermore, the total creation probability declines and the OAM spectrum width increase at larger photon TAM values, as illustrated in Fig. 2(b).

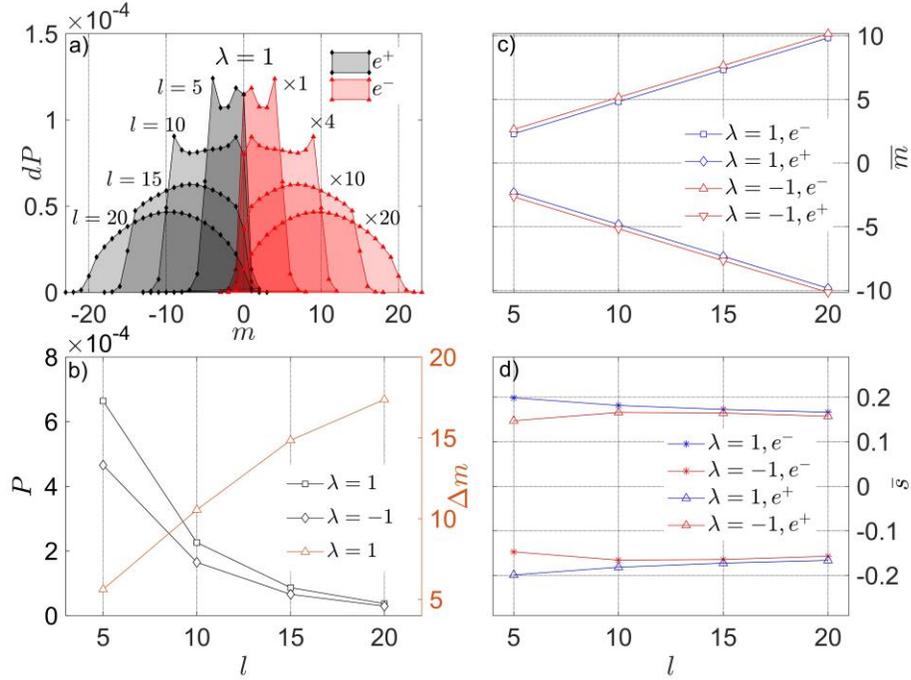

**Figure 2**. The OAM spectra of generated positron (gray) and electron corresponding to the photon TAM number of 5, 10, 15 and 20, respectively (a). The total creation probability and the OAM spectrum width as a function of the photon TAM (b). The average OAM and spin value of the electron-positron pair (c) & (d) in the cases when photon TAM is parallel to polarization ($\lambda$=1) and anti-parallel ($\lambda$=-1), respectively.

Fig. 2(c) shows the averaged OAM number of created electron and positron, after summing over the lepton spins. It is seen that electron and positron gain equal average

OAM, which are $\bar{m}_1 = -\bar{m}_2 = 2.302(2.646)$, $4.820(5.165)$, $7.330(7.666)$, $9.836(10.160)$ for $\lambda=1(-1)$ with $l=5, 10, 15, 20$, respectively and the corresponding equal average SAM are $\bar{s}_1 = -\bar{s}_2 = 0.199(-0.147), 0.181(-0.166), 0.172(-0.164), 0.166(-0.157)$. In either case the symmetry among the electron and positron is preserved $\bar{m}_1 + \bar{s}_1 = -\bar{m}_2 - \bar{s}_2 = l/2$ and the partition of OAM satisfies the average TAM conservation: $\bar{m}_1 - \bar{m}_2 + \bar{s}_1 - \bar{s}_2 = l$.

Flip of the photon polarization also changes the orientation of lepton spins. It is seen in Fig. 2(d) that the averaged spin number turns over when switching from $\lambda=1$ to -1 which leads to higher average OAM in Fig.2(c) (red line). In fact, the photon spin effect is also imprinted in the asymmetric OAM distributions in Fig. 2(a) at relatively small photon TAM numbers, e.g., $l=5$ and 10. This effect becomes less significant when $l$ is large. We show the total scattering probability as a function of photon TAM number in Fig. 2(c) with different photon spins. It is seen that at each TAM value the probability is notably higher for $\lambda=1$. The difference is much more suppressed when increasing the TAM.

To reveal how photon's polarization affects the distribution of positron and electron OAM, we divide the interaction into four channels determined by the final spin states of the leptons and compare their OAM spectra in Fig. 3 with photon TAM $l=10$ and 20 respectively. First of all, for $\lambda=1$ we find that the (-1/2, -1/2) channel in Fig. 3(b) & (f) is significantly suppressed as compared to the (1/2, 1/2) one in Fig. 3(c) & (g). The situation is reversed when switching to $\lambda=-1$ case. The trend presented here are consistent with polarized scattering of plane-wave states. Channel (b) (-1/2, -1/2) and (c) (1/2, 1/2) are independently symmetrical, exhibiting centers at $m_2 = -5.5$ and $m_2 = -4.5$ ($l=10$, $\lambda=1$). This is related to the $\delta$ functions in Eq. (5) responsible for channel (b)

$\delta_{l,m_1-m_2-1}$ and channel (c) $\delta_{l,m_1-m_2+1}$, where there is one unit shift for both the positron and electron. We see that the distribution centers do not necessarily correspond to the spectrum peaks. In fact, the two probability peaks are located at $m_2 = -10$ and -1 in Fig. 2(b), corresponding to $m_1 = 1$ and 10 for the electron (not shown here), respectivly. Therefore the combinations of electron-positron OAM (1, -10) and (10, -1) dominate the channel. Adding the other channels lead to the non-monopole profiles in Fig. 2(a) at $l$=10.

Flipping the photon spin to $\lambda = -1$ induces large variation to the profiles of channel (a) and (d), showing enormous shift of the peak positrons. The distributions indicate that $(m_1, m_2) = (9, -1)$ and $(1, -9)$ are primary. We sum the two distributions (both have $s_1 + s_2 = 0$) up to produce symmetrical spectra centered at $m_2 = -5$ for each photon spin state. As a result, the major difference in terms of average OAM stems from change of spectrum center from channel (c) ($\lambda = 1$) to channel (b) ($\lambda = -1$). This is why the leptons gain slightly larger OAM in the latter case.

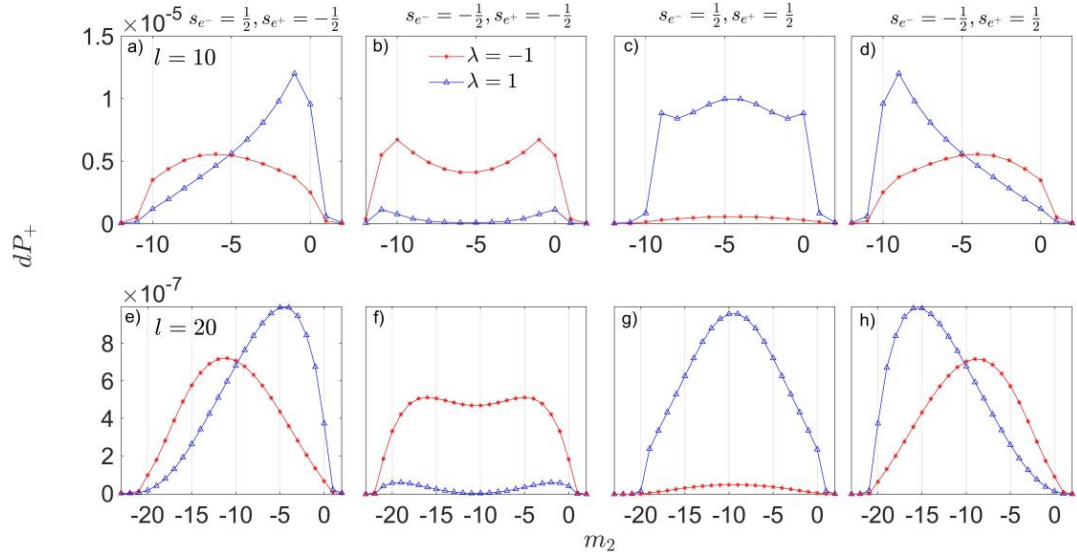

**Figure 3.** Four channels of the OAM spectrum for positron when the photon polarization is varied between $\lambda = 1$ (blue) and $\lambda = -1$ (red). Here photon TAM is $l = 10$ (a-d) and 20 (e-h), respectively.

For *l*=20 the spectra of Fig. 3(f) and (g) much more smooth and monopole-like. After summing up the four channels, the OAM distribution shows good symmetry as seen in Fig. 2(a). This is because the ratio between photon spin and TAM is much smaller here. As mentioned above, large disparity of the scattering probability between different photon spins is found in Fig. 2(b). We notice that at *l*=10 the quantity related to the photon OAM is approximately *l*-*λ*=9 for *λ*=1 and *l*-*λ*=11 for *λ*=-1. To make a comparison at commensuate OAM value, one should choose *l*=8 for *λ*=-1 rather than *l*=10. We have the scattering probabilities of $P = 2.26 \times 10^{-4}$ with (*l*=10, *λ*=1, *l*-*λ*=9) and $P = 2.4 \times 10^{-4}$ with (*l*=8, *λ*=-1, *l*-*λ*=9). In other words, the difference with the same *l*-*λ* vanishes. The obervation suggests that the photon spin has little effect on the total probability when keeping *l*-*λ* (rather than *l*) constant.

## IV. Discussion

Spin-orbit coupling naturally exists in the quantum vortex state in the relativistic regime. The average OAM and spin angular momentum (SAM) of the vortex positron state after considering the SOI are [18]:

$$\langle m \rangle = \hbar (m + \Delta s) \hat{z}, \qquad \langle s \rangle = \hbar (s - \Delta s) \hat{z}. \tag{10}$$

Here $\Delta = (1 - M/E) \sin^2 \theta$ denotes the SOI induced change. We take photon TAM $l = 5$ with different polarization to calculate the possibility as a function of the average OAM and SAM in Fig. 4. As the opening angle $\theta$ increases, both the average OAM $\langle m \rangle$ and SAM $\langle s \rangle$ deviate from the quantized integrals. In general, peak position of average SAM is slightly lower than $|s| = 1/2$ in all cases and the one for $\langle m \rangle$ is also quite close to the OAM quantum number. These results validates our approximations in the analysis.

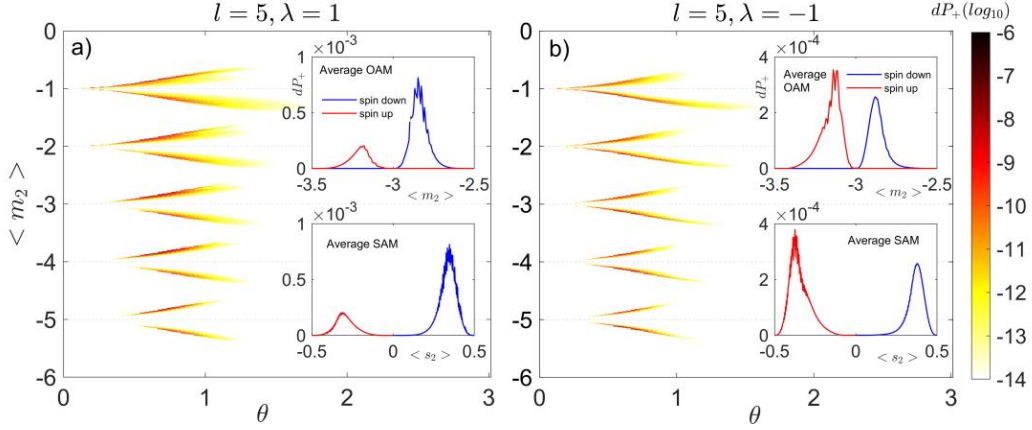

**Figure 4.** Dependence of the scattering probability on the average OAM and opening angle of positron with $l=5$, $\lambda=1$ (a) and $l=5$, $\lambda=-1$ (b) when considering the spin-orbital coupling during interactions. Inner figures show the average OAM and spin after integrating the opening angle, indicating the deviation from the quantum numbers $m_2=-3$ and $s_2=\pm 1/2$.

## V. Conclusions

We employed the twisted states of photons and electrons/positrons to calculate the scattering probabilities of twisted photons into twisted electron-positron pairs under the influence of the Coulomb field. We found that the photon TAM is efficiently transferred to the final leptons. We discussed the effect of spin-orbit coupling on average OAM and SAM and noticed that angular momentum is also transferred from spin to orbit. This work lays the foundation of generating relativistic vortex positron beams for new physics in a number of twisted scattering processes.

**Acknowledgments**

This work is supported by the National Science Foundation of China (Nos. 11875307 and 11935008), the Strategic Priority Research Program of Chinese Academy of Sciences (Grant No. XDB16010000) and the Ministry of Science and Technology of the People's Republic of China (Grant Nos. 2018YFA0404803).

## APPENDIX

The scattering matrix of this process is:

$$M_{fi} = -\frac{e^2}{2}\int d^4x d^4x' \frac{d^4q}{(2\pi)^4} \bar{u}_{p_{1\perp},p_{1z}}^{m_1,s_1}(x) e^{-ip_{1z}z+iE_1 t} \gamma^\mu A_{k_{1\perp},k_{1z},\mu}^{l_1,\lambda_1}(x) \frac{i(\not{q}+M)}{(q^2-M^2)} e^{-iq\cdot(x-x')} \gamma^0 v_{p_{2\perp},p_{2z}}^{m_2,s_2}(r',\theta') e^{-ip_{2z}z'+iE_2 t'} A_0^{Coul}(x')$$

$$-\frac{e^2}{2}\int d^4x d^4x' \frac{d^4q}{(2\pi)^4} \bar{u}_{p_{1\perp},p_{1z}}^{m_1,s_1}(x) e^{-ip_{1z}z+iE_1 t} \gamma^0 A_0^{Coul}(x) \frac{i(\not{q}+M)}{(q^2-M^2)} e^{-iq\cdot(x-x')} \gamma^\nu v_{p_{2\perp},p_{2z}}^{m_2,s_2}(r',\theta') e^{-ip_{2z}z'+iE_2 t'} A_{k_{1\perp},k_{1z},\nu}^{l_1,\lambda_1}(x')$$

(A1)

Calculate the two items separately, the first item (4) becomes:

$$\mathcal{M}_{1,fi} = -\frac{iZe^3}{4(2\pi)} \sqrt{\frac{(E_1-M)(E_2-M)}{\omega E_1 E_2}} \frac{1}{|p_1||p_2|} \delta(\omega - E_1 - E_2)$$

$$\times \xi^{s_1 \dagger} \left( I_1 + II_1 + III_1 + IV_1 + V_1 + VI_1 + VII_1 + VIII_1 + VIIII_1 \right) \eta^{s_2} \Big|_{\substack{E_q = -E_2 = E_1 - \omega \\ q_z = p_{1z} - k_{1z}}}$$

(A2)

$I_1 \sim VIIII_1$ are all $2\times 2$ matrices, through integration, we find that the items at the corresponding positions have the same delta function. Take $I_1$ as an example, the ultimate result is:

$$I_1 = p_{1z} p_{2z}(E_q + M) \int \frac{dq'_\perp q'_\perp}{(p_{1z} + p_{2z} - k_{1z})^2 + q'^2_\perp} \frac{dq_\perp q_\perp}{\left(E_q^2 - (p_{1z} - k_{1z})^2 - q_\perp^2 - M^2\right)}$$

$$\times \begin{pmatrix} \frac{\lambda_1 k_{1\perp}}{\omega_1} S_{l_1}^{m_1}(k_{1\perp},p_{1\perp},q_\perp) S_{m_2}^{m_2}(q_\perp,p_{2\perp},q'_\perp) \delta_{m_2,m_1-l_1} & -i\left(1+\frac{\lambda_1 k_{1z}}{\omega_1}\right) S_{l_1-1}^{m_1}(k_{1\perp},p_{1\perp},q_\perp) S_{m_2}^{m_2}(q_\perp,p_{2\perp},q'_\perp) \delta_{m_2,m_1-l_1+1} \\ -i\left(1-\frac{\lambda_1 k_{1z}}{\omega_1}\right) S_{l_1+1}^{m_1}(k_{1\perp},p_{1\perp},q_\perp) S_{m_2}^{m_2}(q_\perp,p_{2\perp},q'_\perp) \delta_{m_2,m_1-l_1-1} & -\frac{\lambda_1 k_{1\perp}}{\omega_1} S_{l_1}^{m_1}(k_{1\perp},p_{1\perp},q_\perp) S_{m_2}^{m_2}(q_\perp,p_{2\perp},q'_\perp) \delta_{m_2,m_1-l_1} \end{pmatrix}$$

(A3)

Among them we define the integral of the triple-Bessel product:

$$S_n^m(p,k,q) = \int dr r J_n(pr) J_{m-n}(qr) J_m(kr) = (-1)^m \int dr r J_n(pr) J_{m-n}(qr) J_{-m}(kr)$$

$$= \frac{(-1)^m \delta}{2\pi A_{p,q,k}} \cos\left(n(\pi - \angle_{p,q}) + m(\pi - \angle_{k,q})\right) = \frac{(-1)^n \delta}{2\pi A_{p,q,k}} \cos\left(n\angle_{p,q} + m\angle_{k,q}\right)$$

$$= \frac{(-1)^{m+n} \delta}{2\pi A_{p,q,k}} \cos\left(m\angle_{p,k} + (m-n)\angle_{p,q}\right) = \frac{\delta}{2\pi A_{p,q,k}} \cos\left(n\angle_{p,k} - (m-n)\angle_{k,q}\right)$$

(A4)

$I_1 \sim VIIII_1$ are similar to $I_1$, each matrix element is the product of two $S_n^m(p,k,q)$ functions (with different m and n coefficients) and multiplied by the corresponding coefficients that depend on the photon momentum energy and polarization.

$\mathcal{M}_{2,fi}$ is calculated in the same way, and finally we can get

$$\mathcal{M}_{fi} = \mathcal{M}_{1,fi} + \mathcal{M}_{2,fi}$$
$$= \frac{iZe^3}{4(2\pi)}\sqrt{\frac{(E_1-M)(E_2-M)}{\omega E_1 E_2}} \frac{1}{|\boldsymbol{p}_1||\boldsymbol{p}_2|}\delta(\omega-E_1-E_2)\xi^{s_1 \dagger}(\Xi\Big|_{\substack{E_q=-E_2=E_1-\omega \\ q_z=p_{1z}-k_{1z}}} + \tilde{\Xi}\Big|_{\substack{E_q=E_1=\omega-E_2 \\ q_z=k_{1z}-p_{2z}}})\eta^{s_2}. \quad (A5)$$